\pdfoutput=1
\documentclass[12pt]{article}
\usepackage[utf8]{inputenc}
\usepackage{authblk}
\usepackage{amssymb}
\usepackage{graphicx}
\usepackage[margin=1cm]{caption}
\usepackage{geometry}
\usepackage{setspace}

\usepackage{tikz}
\usetikzlibrary{shapes,arrows}
\usepackage{enumitem}

\providecommand{\keywords}[1]
{
  \small	
  \textbf{\textit{Keywords }} #1
}

\providecommand{\version}[1]
{
  \small	
  \textbf{\textit{Git revision }} #1
}

\title{A new multivariate meta-analysis model for many variates and few studies}
\author[1,*]{CJ~Rose}
\author[2,3]{U~Olsen}
\author[2,3]{MF~Lindberg}
\author[1]{E~M-L~Denison}
\author[3]{A~Aamodt}
\author[4,5]{A~Lerdal}

\affil[1]{Division for Health Services, 
          Norwegian Institute of Public Health, 
          Oslo, Norway}
\affil[2]{Department of Nursing Science,
          Institute of Health \& Society,
          Faculty of Medicine,
          University of Oslo,
          Norway}
\affil[3]{Department of Orthopaedic Surgery,
          Lovisenberg Diaconal Hospital,
          Oslo, Norway}
\affil[4]{Department of Patient Safety \& Research,
          Lovisenberg Diaconal Hospital,
          Oslo, Norway}
\affil[5]{Department of Interdisciplinary Health Sciences,
          Institute of Health \& Society,
          Faculty of Medicine,
          University of Oslo, Norway}
\affil[*]{Corresponding author}

\date{\small\today}

\begin{document}

\maketitle

\begin{abstract}
Studies often estimate associations between an outcome and multiple variates.
For example, studies of diagnostic test accuracy estimate sensitivity and
specificity, and studies of predictive and prognostic factors typically estimate
associations for multiple factors. Meta-analysis is a family of statistical
methods for synthesizing estimates across multiple studies. Multivariate models
exist that account for within-study correlations and between-study
heterogeneity. The number of parameters that must be estimated in existing
models is quadratic in the number of variates (e.g., risk factors). This means
they may not be usable if data are sparse with many variates and few studies. We
propose a new model that addresses this problem by approximating a
variance-covariance matrix that models within-study correlation and
between-study heterogeneity in a low-dimensional space using random projection.
The number of parameters that must be estimated in this model scales linearly in
the number of variates and quadratically in the dimension of the approximating
space, making estimation more tractable. We performed a simulation study to
compare coverage, bias, and precision of estimates made using the proposed model
to those from univariate meta-analyses. We demonstrate the method using
data from an ongoing systematic review on predictors of pain and function after
total knee arthroplasty. Finally, we suggest a decision tool to help analysts
choose among available models.
\end{abstract}


\version{
72ca65b}

\pagebreak

\section*{Introduction}

Meta-analysis is a family of statistical methods used to synthesize estimates of
one or more common variates reported by multiple studies \cite{borenstein_2011}.
The aim is to obtain a single estimate that statistically characterizes the
totality of the available evidence, often including any between-study
heterogeneity. For example, a variate of interest might be the prevalence of a
particular disease, or a risk ratio comparing a treatment to a comparator. As in
the examples, the most commonly-used meta-analysis models are univariate, which
means that each primary study contributes an estimate of a single variate.
Univariate meta-analysis is not necessarily appropriate if there are multiple
variates. While it may be tempting to apply univariate meta-analysis to each
variate separately, this does not account for possible correlations between
variates and does not allow ``borrowing of strength'' across variates and
studies \cite{deeks_2001}. Univariate meta-analyses applied in the multivariate
setting are expected to provide excessively biased and imprecise estimates
\cite{riley_2009}.

The alternative is multivariate meta-analysis, which in principle can model all
variates of interest, as well as any within-study correlation and between-study
heterogeneity, simultaneously. Perhaps the most well-known application of
multivariate meta-analysis within biomedical research is in studying diagnostic
test accuracy (DTA), in which the variates sensitivity and specificity are of
interest \cite{irwig_1995}. It is generally recognized that univariate
meta-analysis is inappropriate for DTA because changing the threshold that
distinguishes positive from negative test results to increase sensitivity will
typically decrease specificity. This correlation between sensitivity and
specificity is not modeled by univariate meta-analyses. Multivariate
meta-analysis is also of use in the study of predictive and prognostic factors,
in which a given outcome may be associated with more than one factor. Network
meta‐analysis (multiple treatment comparison) can also be posed as multivariate
meta-analysis \cite{white_2015}. In addition to the challenges faced in
univariate meta-analysis, the multivariate setting poses additional ones, some
or all of which may be addressed by available methods:
\begin{enumerate}
  \item It is rare for authors of primary studies to report an estimate of the
        full variance-covariance or correlation matrix.
  \item It cannot be assumed that every primary study provides estimates for all
        variates of interest.
  \item As in univariate meta‐analysis, it is typical to observe between-study
        heterogeneity in the estimates.
\end{enumerate}

A further challenge, addressed herein, is the scenario in which the number of
variates (e.g., prognostic factors) is large relative to the number of primary
studies. It may not be possible to use existing multivariate meta-analysis
models in such circumstances because the number of parameters needed to estimate
within‐study correlation and between-study heterogeneity is quadratic in the
number of variates. Our contribution is to use a low-dimensional
variance-covariance matrix that approximates common within‐study correlation and
between-study heterogeneity. We do this using a dimensionality reduction method
called random projection \cite{indyk_1998, dasgupta_2000}. This allows us to
reduce the number of parameters that must be estimated to be linear in the
number of variates. Estimation is thereby more tractable when there are few
studies and many variates.

This paper begins with a motivating example from an ongoing systematic review of
predictive factors in which existing methods could not be used. We then provide
mathematical background on multivariate meta-analytical methods and explain in
more detail why estimation is challenging when there are many variates and few
studies. We introduce our model and present a simulation study that compares the
proposed method to univariate meta-analysis with respect to bias, variance, and
coverage probability. We close with a discussion that presents a decision aid
for choosing among the models considered herein, and suggest avenues for
future research.

\section*{Motivating example}

\begin{figure}[b!]
  \centering
  \includegraphics[width=0.6\linewidth]{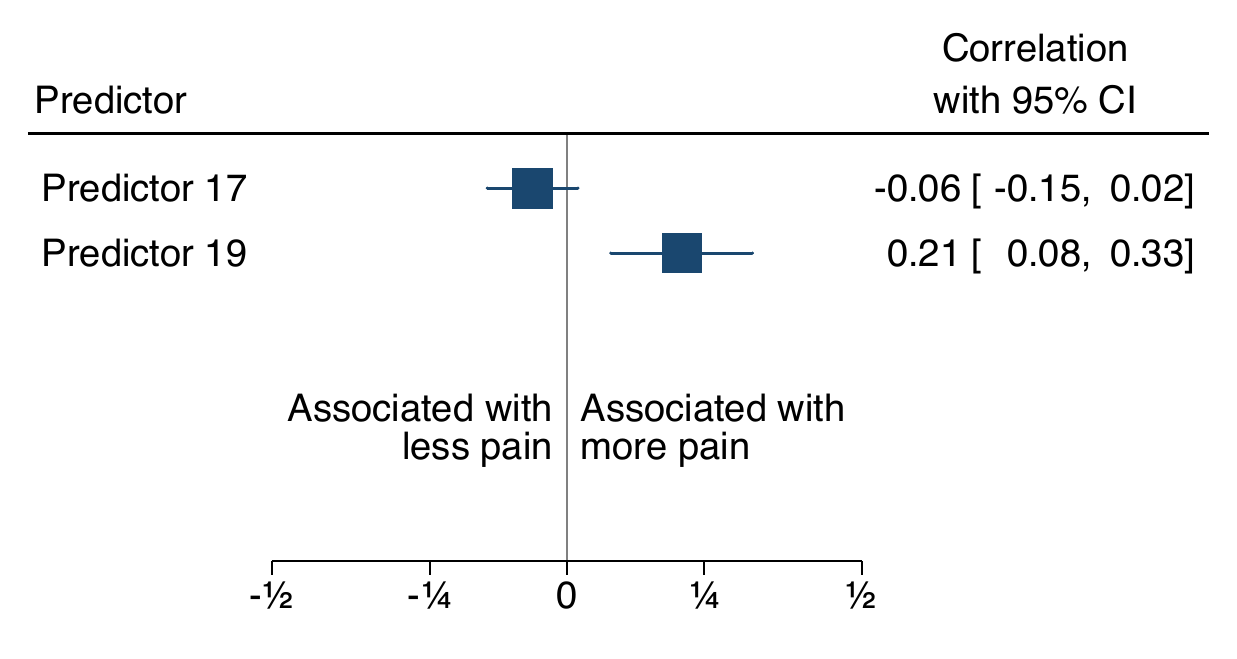}
  \caption{Multivariate meta-analysis could only be performed for two of the 23
           predictors using the model of Riley et al.}
  \label{fig:mvmeta_forest}
\end{figure}

This work was motivated by an ongoing systematic review of factors that may
predict chronic pain and physical function after total knee arthroplasty (TKA)
\cite{olsen_2020}. About 20\% of patients who undergo TKA experience
post-surgical pain and reduced function \cite{beswick_2012}, and numerous
factors have been studied. Being able to characterize factors predictive of
post-surgical pain could lead to better health outcomes and resource use.
Following the inclusion criteria specified in our protocol, we extracted or
imputed 38 estimates of correlation coefficients between 23 predictors and
six-month post-surgical pain (a prespecified secondary outcome) from 7 studies
that included a total of 5473 patients (approximately 2700 patient-years of
follow-up). We had planned to perform multivariate meta-analysis using an
extension of the common correlation model of Riley et~al.~\cite{riley_2008}, as
implemented in the MVMETA add-on command for Stata \cite{white_2009,
white_2011}. However, because the extracted data are sparse, it was not possible
to perform the prespecified analysis unless we applied some criterion to limit
the number of studies and predictors included in the analysis. We found that
including predictors supported by at least four studies allowed the prespecified
model to be used. However, this approach has at least two clear problems. First,
the choice of criterion represents an investigator degree of freedom, and it is
possible that several other reasonable criteria would each allow the model to be
fitted, likely yielding differing results. Second, the criterion we chose
limited the number of predictors for which multivariate meta-analysis estimates
could be obtained to just two (see figure \ref{fig:mvmeta_forest}). Recognizing
that the planned analysis was suboptimal, we then attempted to use Lin and Chu's
model \cite{lin_chu_2018}, which was developed to address the problem of sparse
data in multivariate meta-analysis. Unfortunately, the available data were also
too sparse to support this model.

\section*{Background}

Let $\mathbb{S}$ be a set of $m$ studies. The $i$-th study provides $1 \leq t_i
\leq p$ point estimates and sampling variances, where $p$ is the total number of
unique variates studied\footnote{We use the word ``variate'' synonymously with
``endpoint'' of Riley et~al.~\cite{riley_2008} and ``factor'' of Lin and Chu
\cite{lin_chu_2018}.}. Let the point estimates provided by study $i$ be denoted
$y_i \in \mathbb{R}^{t_i}$ and the corresponding diagonal matrix of sampling
variances be denoted $D_i \in \mathbb{R}_{\geq0}^{t_i \times t_i}$. Given the
$y_i$ and $D_i$, we wish to estimate the true value of the $p$ variates, $\mu
\in \mathbb{R}^p$, accounting for within‐study correlation and between-study
heterogeneity. We assume that none of the studies report within-study
correlation or variance-covariance matrices.

Riley et~al.~\cite{riley_2008} proposed a bivariate meta-analysis model that
assumes a common within-study correlation parameter. A multivariate version of
this model has been implemented for Stata by White \cite{white_2009,
white_2011}. Assuming such a model is parameterized in terms of a common
within-study correlation matrix and a between-study variance-covariance matrix
that models heterogeneity, the parameters to be estimated are the $p$ elements
of $\mu$, the $\frac{1}{2} p (p - 1)$ elements of the common within-study
correlation matrix (which is unitriangular), and the $\frac{1}{2} p (p + 1)$
elements of the upper or lower triangle of the variance-covariance matrix that
models heterogeneity. Such a model requires a total of $p^2 + p$ parameters to
be estimated. It may be challenging to fit such a model unless the total number
of point estimates provided by the studies $n = \sum_{i \in \mathbb{S}} t_i \geq
p^2 + p$. For many research questions, sufficient studies and estimates may not
exist, particularly if $p$ is large. We say the problem is sparse if $n < p^2 +
p$.

Lin and Chu \cite{lin_chu_2018} developed on the model of Riley et
al. and addressed the sparsity problem by modeling the variance-covariance
matrix for each study as a sum of sampling and additional variances, and by
assuming a common correlation matrix for all $p$ variates. Their model requires
estimating the $p$ elements of $\mu$, the $p$ additional variances, and the
$\frac{1}{2} p (p - 1)$ correlations, for a total of $\frac{1}{2} p (p + 3)$
parameters.

\begin{table}
  \centering
  \begin{tabular}{lr}
    \hline
    \textbf{Model}  & \textbf{Number of Parameters} \\
    \hline
    Riley et~al.~   & $p^2 + p$                     \\
    Lin and Chu     & $\frac{1}{2} p (p + 3)$       \\
    Our Model       & $p + \frac{1}{2} q (q - 1)$   \\
    \hline
  \end{tabular}
  \caption{The number of parameters that must be estimated for each model;
           $p$ is the total number of variates across all studies, and $q$
           is the dimensionality of the space in which within-study correlation
           and between-study heterogeneity are modeled.}
  \label{table:parameters}
\end{table}

\section*{A low-dimensional model}

As in previous work, we make the simplifying assumptions that variates not
estimated by particular studies are missing completely at random, and have a
common correlation structure across the primary studies \cite{lin_chu_2018,
riley_2008}. We develop on Lin and Chu's model by assuming that the
within-study correlation structure and between-study heterogeneity are well
approximated in a low-dimensional space. This allows us to reduce the number of
parameters that must be estimated. Our model is:
\begin{equation}
  y_i \sim \mathcal{N}(X_i \mu, \Phi_i) \textrm{ for } i \in \mathbb{S}
\end{equation}
where
\begin{equation}
  \Phi_i = D_i + X_i R^\top \Sigma R X_i^\top
\end{equation}

\begin{figure}[b!]
  \centering
  \includegraphics[width=0.7\linewidth]{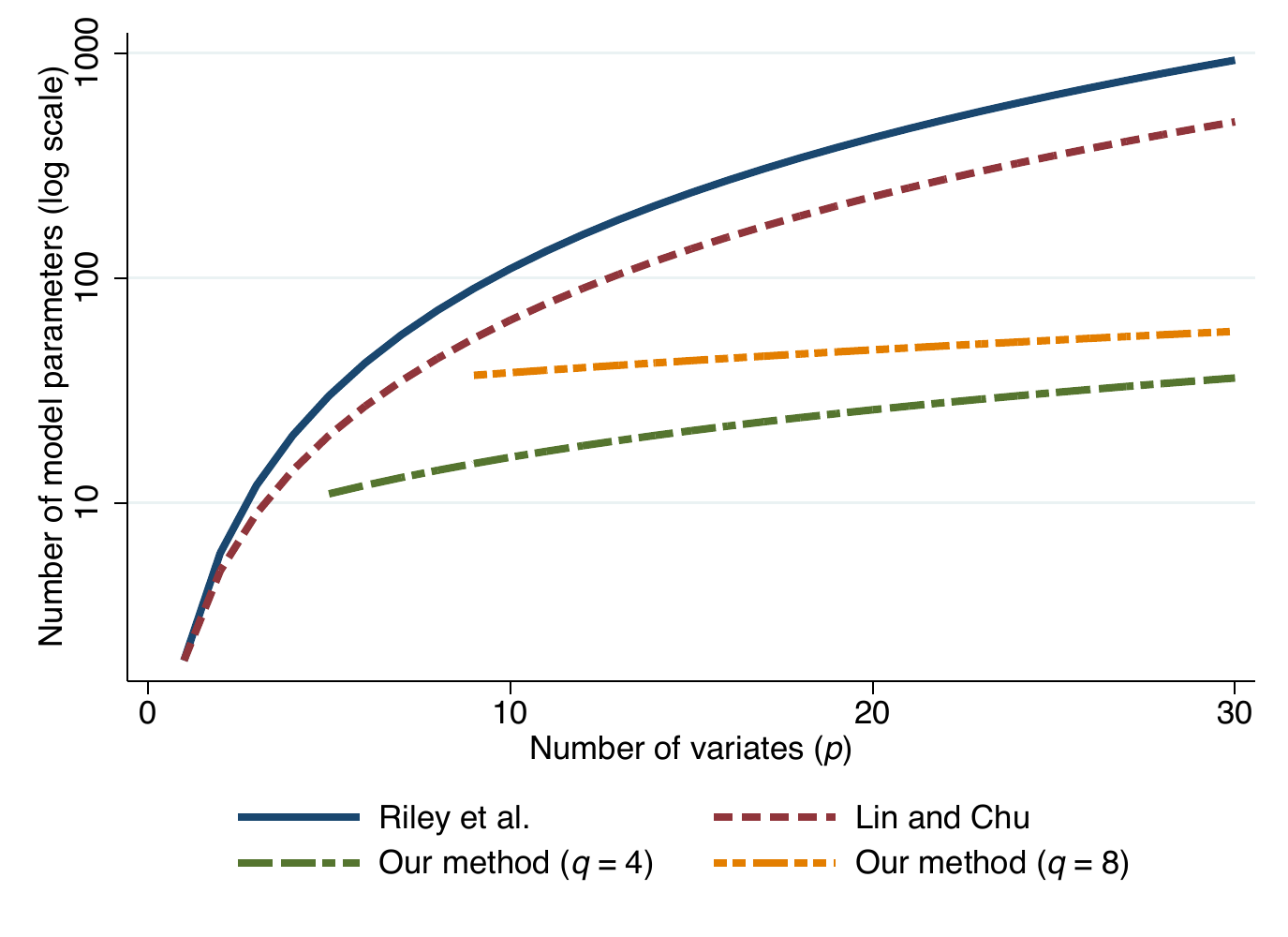}
  \caption{Number of model parameters as a function of number of variates for
           Riley's and Lin and Chu's models, and our model with $q = 4$ and $q =
           8$. Curves for our model are shown for $q < p$.}
  \label{fig:p_vs_num_parameters}
\end{figure}

$X_i$ is a $t_i \times p$ indicator matrix. Its $(j, k)$-th element is unity if
the $j$-th estimate reported by the $i$-th study corresponds to the $k$-th of
the $p$ variates and is zero otherwise. $R \in \mathbb{R}^{q \times p}$ is a
matrix that maps between the full $p$-dimensional space and a $q$-dimensional
space (where $q < p$) in which the full within- and between-study
variance-covariance matrix is approximated by the symmetric positive-definite
matrix $\Sigma \in \mathbb{R}^{q \times q}$. Table
\ref{table:parameters} summarizes the number of parameters that must be
estimated by the models of Riley et al., Lin and Chu, and our model. In brief,
for fixed $q$ (which would reasonably be in the range 2--10), the number of
parameters that must be estimated for our model scales as $O(p)$ rather than
$O(p^2)$, as for those of Riley et~al.~and Lin and Chu. Figure
\ref{fig:p_vs_num_parameters} plots number of model parameters as a function of
number of variates for Riley's and Lin and Chu's models, and our model with $q =
4$ and $q = 8$.

We use a random projection matrix \cite{dasgupta_2000, jl_2984} for $R$, but any
suitable dimensionality-reduction method could be used instead. There is a large
literature on the theory and applications of random projection, and
\cite{dasgupta_2000} provides a good introduction. The following is a brief and
informal treatment of the relevant concepts. Further background on the approach
is provided in the Discussion section.

Recall that an orthogonal linear transformation matrix $T \in \mathbb{R}^{p
\times p}$ maps between orthonormal bases and preserves the magnitudes and
angles between vectors. Such transforms are of interest in multivariate
statistics, in particular in principal component analysis (PCA), which has
application in dimensionality reduction. Briefly, PCA can be posed as follows:
given a variance-covariance matrix $\Lambda$, find matrix $T$ such that
$T\Lambda{}T^\top$ is a diagonal matrix of variances that preserves the total
variance of $\Lambda$. This can be achieved via eigendecomposition of $\Lambda$,
giving a matrix of eigenvectors ($T$) and their associated eigenvalues, the
latter of which provide information about the proportion of the total variance
explained in the direction of each of the eigenvectors. Dimensionality reduction
can be achieved via the matrix $T' \in \mathbb{R}^{q \times p}$, with $q < p$.
$T'$ is formed by dropping those eigenvectors from $T$ that have the smallest
associated eigenvalues. Hence, a dimensionality-reducing transform can be
defined that preserves at least a given proportion of the total variance. In
short, the original matrix $\Lambda$ can be approximated in a $q$-dimensional
space by $T'\Lambda{}T'^\top$.

PCA is only applicable if the variance-covariance matrix is known or can be
estimated. Random projection is a method of establishing an approximately
orthogonal linear transform, $R$, that defines a basis in a low-dimensional
space. Interestingly, the elements of $R \in \mathbb{R}^{q \times p}$ are
iid samples from a particular distribution. A variance-covariance matrix
$\Lambda$ can be approximated in a $q$-dimensional space by $R\Lambda{}R^\top$.
There are three key differences between PCA and random projection that are
relevant to our model:

\begin{enumerate}
  \item Unlike PCA, a random projection matrix can be constructed without
        knowing anything about the variance-covariance matrix, except for its
        dimension.
  \item Unlike PCA, $R\Lambda{}R^\top$ is not necessarily diagonal. It is
        therefore necessary to estimate all elements of the low-dimensional
        matrix $\Sigma$, which is then transformed to the $p$-dimensional space
        by $R^\top \Sigma R$.
  \item The eigenvectors and eigenvalues obtained in PCA are often of interest
        to the analyst in their own right. For example, it may be useful to know
        that 95\% of total variance can be explained by three principal
        components, and how they relate to the original variates. In random
        projection, however, provided the transform defines a sufficiently
        useful basis, it is of little interest to the analyst because it is
        arbitrary.
\end{enumerate}

The number of parameters to be estimated could be reduced further by modeling
$\mu$ in the same $q$-dimensional space via $\beta = R\mu \in \mathbb{R}^q$.
While modeling within- and between-study variances and covariances in a
low-dimensional space might be expected to lead to poorer quantification of
precision, modeling $\mu$ in this way might be expected to lead to bias, which
is arguably more serious.

\section*{Simulation study}

\begin{figure}[t!]
  \centering
  \resizebox{0.8\columnwidth}{!}{%
    \sffamily
    \footnotesize
      \begin{tikzpicture}[auto,
        ampersand replacement=\&,
        line/.style = {draw, thick, -latex', shorten >=0pt},
        notext/.style = {rectangle, node contents =,},
        wider/.style ={rectangle, draw=black, thick, fill=white,
          text width=16em, text ragged, minimum height=3em, inner sep=6pt},
        block/.style ={rectangle, draw=black, thick, fill=white,
          text width=10em, text ragged, minimum height=3em, inner sep=6pt},
        center/.style ={rectangle, draw=black, thick, fill=white,
          text width=10em, text centered, minimum height=3em, inner sep=6pt}]
        \matrix [column sep=3mm,row sep=10mm] {
          {} \&
          \node [wider] (sim) {Simulated meta-analyses (n=1\,000)};\\
          \node [center] (uni) {Analyzed using univariate meta-analysis (n=1\,000)}; \&
          {} \&
          \node [center] (mul) {Analyzed using sparse multivariate meta-analysis (n=1\,000)};\\
          {} \&
          \node [block] (exu) {Excluded (n=2):\\
            \begin{itemize}[leftmargin=*, nolistsep]
              \item Did not converge (n=1)
              \item Zero-length CI (n=1)
            \end{itemize}}; \&
          {} \&
          \node [block] (exm) {Excluded (n=6):\\
            \begin{itemize}[leftmargin=*, nolistsep]
              \item Did not converge (n=6)
            \end{itemize}}; \\
          {} \&
          \node [wider] (ana) {Analyzed:\\
            \begin{itemize}[leftmargin=*, nolistsep]
              \item Simulated meta-analyses (n=993)
              \item CI-CrI pairs (n=15\,291)
            \end{itemize}};\\
        };
        \begin{scope}[every path/.style=line]
          \path (sim) |- (uni);
          \path (sim) |- (mul);
          \path (uni) |- (exu);
          \path (uni) |- (ana);
          \path (mul) |- (exm);
          \path (mul) |- (ana);
        \end{scope}
      \end{tikzpicture}%
      } 
      \caption{Outline of the simulation study.}
      \label{fig:sim_flow}
\end{figure}
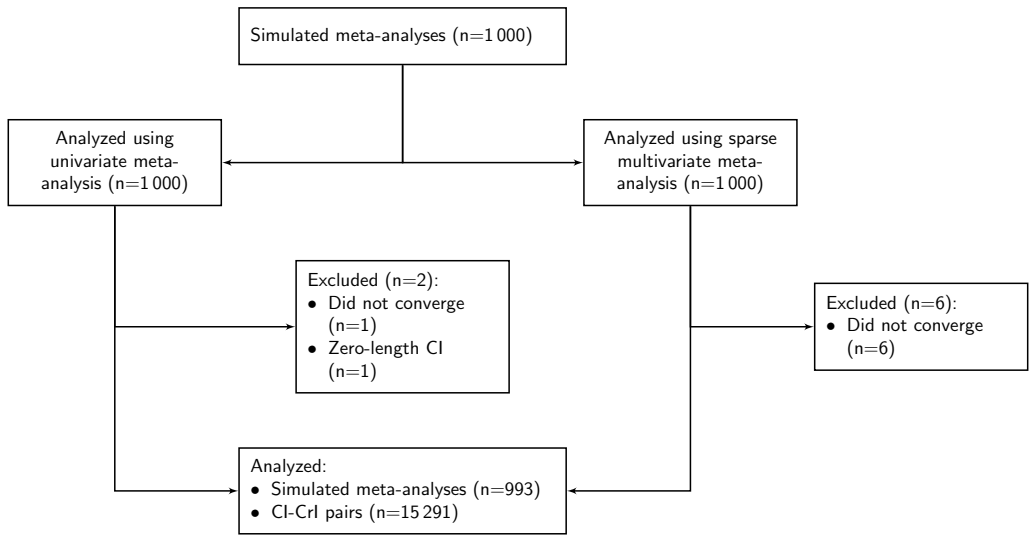

We performed a simulation study to compare the proposed method to univariate
meta-analysis with respect to bias and variance of point estimates, and coverage
probabilities of credible and confidence intervals (CrIs and CIs). Figure
\ref{fig:sim_flow} shows the study flow diagram. We created 1000 random
meta-analysis data sets to be statistically similar to those of the total knee
arthroplasty data as follows. Each simulated meta-analysis data set comprised 4
to 15 simulated primary studies, each contributing point estimates and standard
errors computed from 50 to 4000 simulated units (e.g., patients), with both
numbers sampled uniformly. At the level of meta-analysis, we chose a random
correlation matrix to define target parameter values. At the level of study
within meta-analysis, we sampled from the multivariate normal defined by the
chosen correlation matrix and computed sample correlations to estimate the
target values; one of the dimensions was used to model outcome variable
(analogous to post-surgical pain in the total knee arthroplasty example) and the
others to model the variates (analogous to the risk factors of pain). To
facilitate meta-analysis of the correlations, we applied Fisher's $z$~transform
(hyperbolic arctangent function) and computed the associated standard errors. We
added a normally-distributed value to the estimates to simulate within-study
heterogeneity, with the variance of the distribution chosen to give $I^2$ values
similar to those observed in the knee data. Finally, we assumed that variates
were missing completely at random within each study, creating data sets with
similar density (ratio of total number of estimates to the number that would be
available if all variates were studied) to the knee data.

Each simulated meta-analysis data set was analyzed using univariate
meta-analysis and the method proposed herein. We did not include Riley's or Lin
and Chu's models because our model addresses the scenario in which data are too
sparse for these models to be used. We performed univariate analysis using
Stata's \texttt{meta regress} with variate as a categorical covariate, to fit a
random effects model using restricted maximum likelihood. We implemented our
model within the Bayesian framework using Stan version 2.24.1
\cite{carpenter_2017}, although frequentist implementations would also be
possible. We used the priors $\mu~\sim~\mathcal{N}(0, 10^3)$ and
$\Sigma~\sim~\mathcal{W}^{-1}(I, q + 1)$, where $\mathcal{W}^{-1}$ is the
inverse Wishart distribution and $I$ is the identity matrix. For each simulated
meta‐analysis, we modeled the within- and between-study variance-covariance
matrix using the largest $q$ such that the number of parameters to be estimated
was no greater than the number of estimates available. Estimation was performed
by running four Hamiltonian Monte Carlo chains concurrently using GNU parallel
\cite{tange_2020}, sampling using the No-U-Turn Sampler (NUTS)
\cite{hoffman_2014} using default settings. Specifically, we discarded the first
1000 samples from each chain and accepted the subsequent 1000 samples (if
$\hat{R} < 1.01$ for all variates) for a total of 4000 MCMC samples. Exploratory
work showed that using much larger numbers of MCMC samples would give almost
identical results at the cost of substantially more computation. In addition to
the estimates, we recorded which model fits converged, and which provided
zero-width CIs or CrIs. Only usable estimates were included in further analysis.

To compare bias and variance between the two methods, we analyzed relative
absolute bias and relative lengths of CIs and CrIs on the log scale. Neglecting
indices identifying simulated meta-analysis and variate for ease of exposition,
log relative absolute bias was computed as $\log~\|\mu~-~\hat{\mu}_m\|~-~\log
\|\mu~-~\hat{\mu}_u\|$, where $\mu$ is the value of the target parameter, and
$\hat{\mu}_m$ and $\hat{\mu}_u$ are the multi- and univariate estimates,
respectively. Similarly, log relative lengths of CIs and CrIs were computed as
$\log\,(a_m~-~b_m)~-~\log\,(a_u~-~b_u)$, where $a$ and $b$ indicate bounds on
the intervals from above and below, respectively, and the subscripts indicate
method, as before. CIs and CrIs have very different interpretations. Of
relevance here, a 95\% CrI would not necessarily be expected to be of comparable
length to a 95\% CI, nor have 95\% frequentist coverage. A full exposition is
outside the scope of this paper, but in brief CrIs are a function of the model,
data, priors, estimation procedure, and method of construction (e.g.,
equal-tailed versus highest posterior density). Exploratory work suggested that
equal-tailed 95\% CrIs provide approximately 81\% coverage, and that
equal-tailed 98\% CrIs provide approximately 95\% coverage. We therefore use
equal-tailed 98\% CrIs to permit direct comparison to the 95\% CIs provided by
the univariate method.

We estimated coverage, mean log relative bias, and mean log relative lengths. We
also performed regression analyses to characterize associations with the
total numbers of estimates and variates available in a given meta‐analysis,
adjusting for possible within-meta-analysis clustering. We exponentiated
estimated regression coefficients where appropriate to report comparisons as
relative values.

Usable estimates were provided by 99.9\% (95\%~CI 99.4\% to 100\%) of univariate
versus 99.4\% (95\%~CI 98.7\% to 99.8\%) of multivariate analyses. In total,
estimates from 993 of the 1000 simulated meta-analysis data sets could be
analyzed (corresponding to 15\,291 CI-CrI pairs). We estimate coverage
probabilities of 98.2\% (95\%~CI 98.0\% to 98.4\%) for univariate confidence
intervals versus 94.4\% (95\%~CI 94.1\% to 94.8\%) for multivariate CrIs. There
was no association between coverage and total numbers of estimates and variates
for either method. This suggests that coverage probability does not degrade with
number of variates, for example.

\begin{figure}
  \centering
  \includegraphics[width=\linewidth]{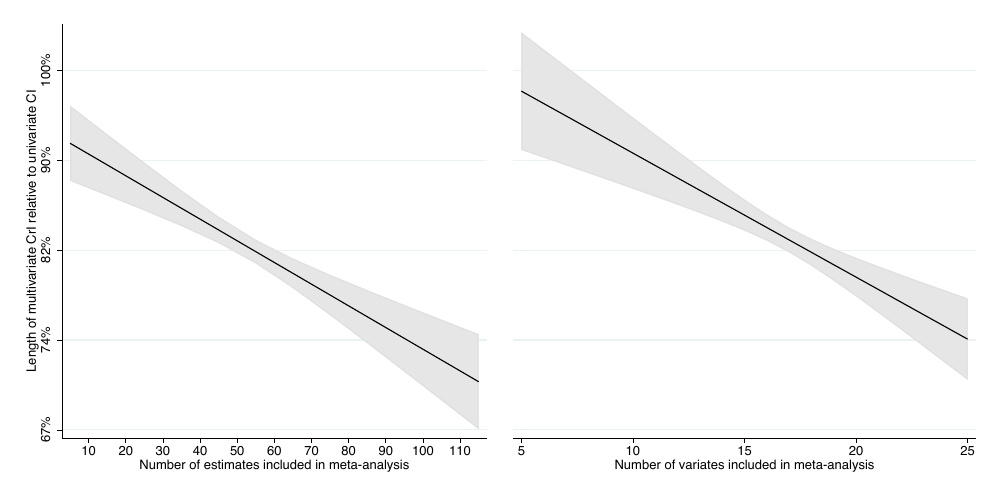}
  \caption{Mean relative lengths (with pointwise 95\% confidence bands) of
           credible intervals provided by the multivariate model compared to
           univariate confidence intervals with similar coverage. Values less
           than 100\% favor the multivariate method.}
  \label{fig:relative_lengths}
\end{figure}

Our method provides estimates with mean bias 1.04 (95\% CI 1.03 to 1.06) times
higher than univariate meta-analysis. Mean absolute bias for univariate
meta-analysis was estimated to be 0.02 units on the hyperbolic arctangent scale,
corresponding to mean absolute bias for our method of 0.0208 (95\% CI 0.0206 to
0.0212) --- i.e., no additional bias to two decimal places. Our method is biased
relative to the univariate method, but the magnitude of the bias may or may not
be of practical importance, depending on context. Our method provides CrIs of
mean length 0.84 (95\% 0.82 to 0.84) times as long as CIs provided by univariate
meta-analysis. In other words, our method provides credible intervals that are
substantially shorter (i.e., more precise) than the univariate method, with
coverage that is very close to desired 95\%. There was no association between
bias and total numbers of estimates or variates. This suggests that if the
additional precision that our method can provide is desirable and the bias is
acceptable, analysts should not be overly concerned about the number of
estimates and variates available.

Figure \ref{fig:relative_lengths} summarizes the regression results with respect
to the lengths of multivariate CrIs relative to univariate CIs. We estimate that
the relative length of CrIs provided by our method decreases (i.e., improves)
with increasing total numbers of estimates and variates. We estimate that adding
10 estimates to a meta-analysis would result in CrIs that are on average 0.98
(95\% CI 0.97 to 0.98) times as long as CIs provided by univariate
meta-analysis. This may not be important in practice. However, we estimate that
adding 10 variates to a meta-analysis would result in CrIs that are on average
0.87 (95\% CI 0.83 to 0.92) times as long as CIs provided by univariate
meta-analysis. This demonstrates that our model can make effective use of the
additional information provided by additional variates and that analysts should
not be overly concerned about dimensionality.

\section*{Application to knee pain data}

We used our model to analyze the TKA knee pain data introduced in the motivating
example above. As for that analysis, we extracted correlation coefficients or
imputed them from extracted estimates of regression coefficients, risk
ratios, and odds ratios. We converted correlation coefficients using Fisher's
$z$-transform (hyperbolic arctangent function) prior to meta-analysis. Further
details are given in our protocol \cite{olsen_2020}.

Estimation was performed as in the simulation study, with the following
exceptions. We used and discarded the first 10,000 samples from each chain as
burn-in and accepted the next 25,000 for a total of 100,000 posterior samples.
Meta-analytical estimates were transformed back to the correlation coefficient
scale using the hyperbolic tangent function. We computed surface under the
cumulative ranking curve (SUCRA) values \cite{salanti_2011} to quantify the
extent of evidence that the magnitude of the correlation coefficient of each
predictor is greater than for all other predictors, and hence provide an
indicative ranking of predictors. We summarized between-study heterogeneity
using univariate $I^2$ statistics.

\begin{figure}[b!]
  \centering
  \includegraphics[width=0.8\columnwidth]{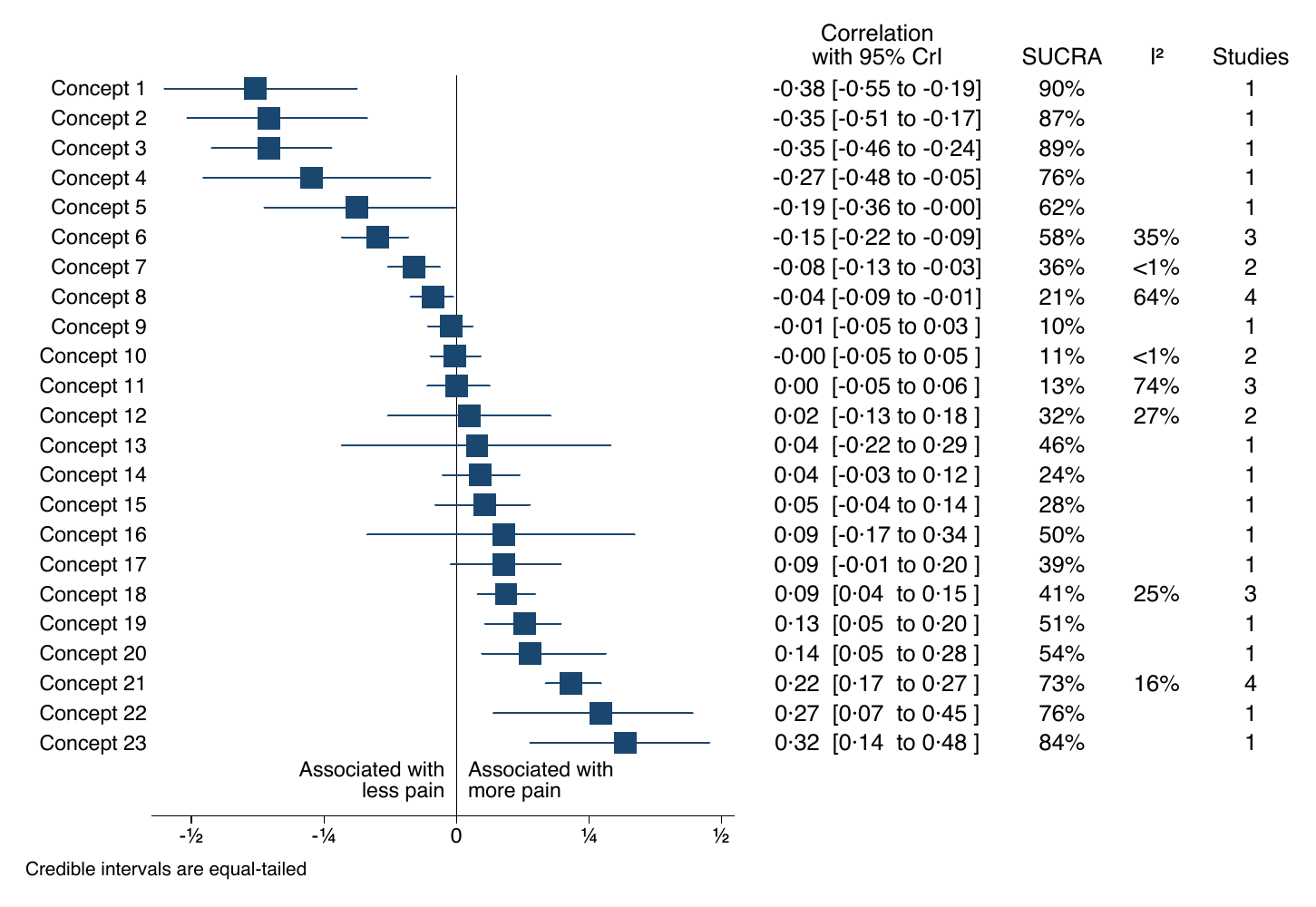}
  \caption{Forest plot showing posterior means and 95\% credible intervals for
           the 23 predictors included in our ongoing systematic review on pain
           and function after total knee arthroplasty.}
  \label{fig:sparse_forest}
\end{figure}

\begin{figure}
  \centering
  \includegraphics[width=0.6\linewidth]{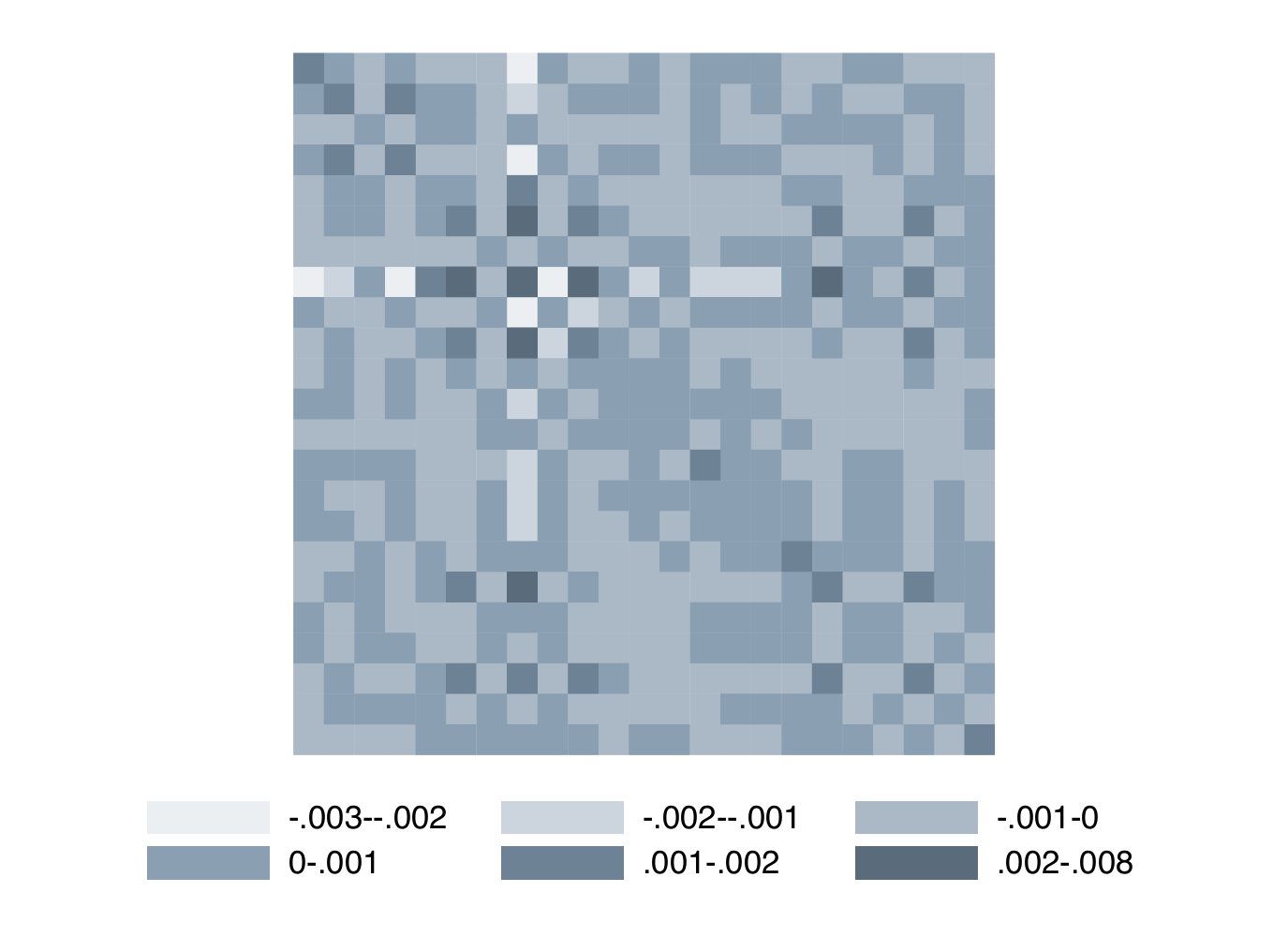}
  \caption{Posterior mean variance-covariance matrix for the total knee
           arthroplasty data.}
  \label{fig:covmat}
\end{figure}

Figure \ref{fig:sparse_forest} shows estimates of correlation coefficients
between each predictor and pain measured six months post-TKA. We present
posterior means and equal-tailed 95\% CrIs. Note that because the
systematic review is ongoing, we have disguised the names of the predictors. We
compared the estimates from our model with those from White's implementation of
the model of Riley et al., and to univariate meta-analysis to assess
consistency. Figure \ref{fig:covmat} shows the posterior mean
variance-covariance matrix $R^\top \Sigma R$ (computed element-wise) estimated
for the TKA knee pain data.

\section*{Discussion}

We have described the problem of multivariate meta-analysis when data are sparse
and proposed a tractable model that approximates within-study correlations and
between-study heterogeneity in a $q$-dimensional space, where $q$ is smaller
than $p$, the total number of variates of interest. The main advantages of this
model are that it can be used when data are too sparse for methods proposed by
Riley et al. and Lin and Chu, that it provides estimates that are on average
substantially more precise compared to univariate meta-analysis (i.e., our
method is statistically more powerful), and that our method provides
increasingly precise estimates as the number of estimates or variates increases.

We model within‐study correlation and between-study heterogeneity in a
low-dimensional space via random projection. Analogously to the preservation of
magnitudes and angles in orthogonal linear transforms, Johnson and Lindenstrauss
showed that it is possible to embed high-dimensional data into much
lower-dimensional spaces while preserving the relative distances between points
\cite{jl_2984}. Indyk et~al.~subsequently proposed using random matrices
\cite{indyk_1998}. A wide range of distributions have been shown to yield good
results with high probability \cite{dasgupta_2000}. While random projection is
typically used in problems in which $p$ is of much higher dimension than is
commonplace in multivariate meta-analysis, Lu and Lió studied random projection
in low dimensions \cite{lu_2008} and showed, perhaps unsurprisingly, that the
distortion introduced when $q$ is small can be negligible if the intrinsic
dimensionality of the original space is low. Our simulation study shows that
our method performs well even though we use random projection in relatively low
dimensions, and that precision improves as $p$ and hence $q$ increases. Random
projection and related methods have subsequently become an important tool in
high-dimensional statistics and machine learning, and have been applied to a
range of problems, including regression \cite{thanei_2017}, mixture modeling
\cite{dasgupta_2000}, text analysis \cite{bingham_2001}, and medical imaging
\cite{lustig_2008}.

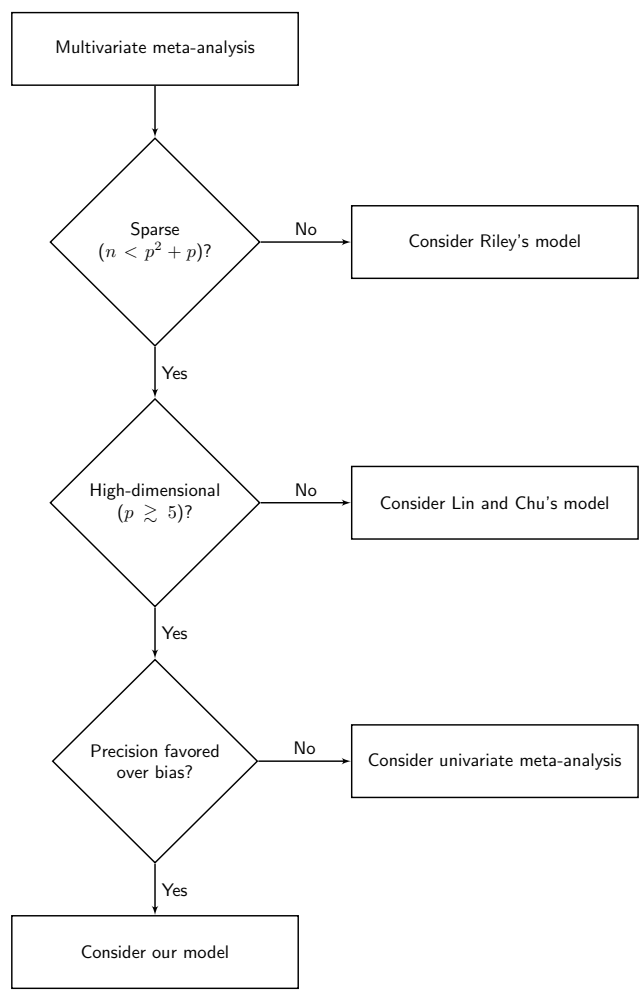
\begin{figure}[t!]
  \centering
  \resizebox{0.5\columnwidth}{!}{%
    \sffamily
    \footnotesize
    \begin{tikzpicture}[auto,
      ampersand replacement=\&,
      block_center/.style ={rectangle, draw=black, thick, fill=white,
        text width=15em, text centered,
        minimum height=4em},
      block_decide/.style ={diamond, draw=black, thick, fill=white,
        text width=8em, text centered,
        minimum height=4em},
      line/.style ={draw, thick, -latex', shorten >=0pt}]
      \matrix [column sep=10mm,row sep=10mm] {
        \node [block_center] (ma) {Multivariate meta-analysis}; \\
        \node [block_decide] (sparse) {Sparse ($n < p^2 + p$)?}; \&
        \node [block_center] (riley) {Consider Riley's model}; \\
        \node [block_decide] (hi_dim) {High-dimensional ($p \gtrsim 5$)?}; \&
        \node [block_center] (lin_chu) {Consider Lin and Chu's model}; \\
        \node [block_decide] (bias_var) {Precision favored over bias?}; \&
        \node [block_center] (univariate) {Consider univariate meta-analysis}; \\
        \node [block_center] (our_method) {Consider our model}; \\
        \\
      };
      \begin{scope}[every path/.style=line]
        \path (ma)       --                           (sparse);
        \path (sparse)   -- node[pos=0.5,above] {No}  (riley);
        \path (sparse)   -- node[pos=0.5,right] {Yes} (hi_dim);
        \path (hi_dim)   -- node[pos=0.5,above] {No}  (lin_chu);
        \path (hi_dim)   -- node[pos=0.5,right] {Yes} (bias_var);
        \path (bias_var) -- node[pos=0.5,above] {No}  (univariate);
        \path (bias_var) -- node[pos=0.5,right] {Yes} (our_method);
      \end{scope}
      \end{tikzpicture}
      } 
    \caption{Suggested approach for choosing a method to meta-analyze
             multivariate data.}
    \label{fig:decision}
\end{figure}

The main disadvantages of our model are that it provides estimates that are more
biased compared to univariate meta-analysis (although the magnitude of the bias
may not be of practical importance, depending on context), the requirement to
choose $q$, and the model’s inability to disentangle the within- and
between-study variances and covariances, which may be of interest in their own
right. However, these disadvantages may be acceptable when $n < \frac{1}{2} p (p
+ 3)$ and it is desirable to obtain more precise estimates than univariate
meta-analysis can provide. Figure \ref{fig:decision} suggests a decision tool
that may help analysts choose an appropriate multivariate meta-analysis model.
However like any rule of thumb, it should not be used unless one understands
when and why one should deviate from the rule, and it does not aim to include
all multivariate meta-analysis methods available.

We suggest that authors wishing to use our model report the possible limitations
introduced by the low-dimensional approximation, and compare the estimates
provided by our model to those from a conventional multivariate meta-analysis
that includes as many variates as possible, as well as to univariate
meta-analyses. Inconsistencies between models should be reported and their
implications explained in a way that can be understood by non-quantitative
decision-makers.

Future work could address estimating the free parameter $q$ (e.g., by placing a
prior over the dimensionality of the model or otherwise integrating over $q$);
using alternative dimensionality-reduction methods; modeling the distinction
between studies that adjusted estimates for other variates, versus studies that
did not; and modeling the scenario in which variates are not missing completely
at random. Estimation in the face of sparse data remains an interesting and
potentially rewarding research area.

\bibliography{ms}

\begin{thebibliography}{2}

\bibitem{beswick_2012}
  Beswick AD, Wylde V, Gooberman-Hill R, et~al.~What proportion of patients
  report long-term pain after total hip or knee replacement for osteoarthritis?
  A systematic review of prospective studies in unselected patients. BMJ Open
  2012;2:e000435

\bibitem{borenstein_2011}
  Borenstein, M., Hedges, L. V., Higgins, J. P., and Rothstein, H. R. (2011).
  Introduction to meta-analysis. John Wiley \& Sons.

\bibitem{bingham_2001}
  Bingham, E. and Mannila, H. (2001). Random projection in dimensionality
  reduction: applications to image and text data. In Proceedings of the Seventh
  ACM SIGKDD International Conference on Knowledge Discovery and Data Mining
  245-250.

\bibitem{carpenter_2017} 
  Carpenter, B., Gelman, A., Hoffman, M. D., Lee, D., Goodrich, B., Betancourt,
  M. et~al.~(2017). Stan: A probabilistic programming language. Journal of 
  Statistical Software, 76(1).

\bibitem{dasgupta_2000}
  Dasgupta, S. (2000). Experiments with random projection. In Proceedings of the
  Sixteenth Conference on Uncertainty in Artificial Intelligence, 143-151.

\bibitem{deeks_2001}
  Deeks, J. J. (2001). Systematic reviews of evaluations of diagnostic and
  screening tests. BMJ, 323(7305), 157-162.

\bibitem{hoffman_2014}
  Hoffman, M. D. and Gelman, A. (2014). The No-U-Turn sampler: adaptively
  setting path lengths in Hamiltonian Monte Carlo. Journal of Machine Learning
  Research, 15(1), 1593-1623.

\bibitem{irwig_1995}
  Irwig, L., Macaskill, P., Glasziou, P., and Fahey, M. (1995). Meta-analytic
  methods for diagnostic test accuracy. Journal of Clinical Epidemiology, 48(1),
  119-130.

\bibitem{jl_2984}
  Johnson, W. B. and Lindenstrauss, J. (1984). Extensions of Lipschitz mappings
  into a Hilbert space. Contemporary Mathematics, 26(189-206), 1.

\bibitem{lu_2008}
  Lu, Y. E., Lió, P., and Hand, S. (2008, October). On low dimensional random
  projections and similarity search. In Proceedings of the 17th ACM Conference
  on Information and Knowledge Management, 749-758.

\bibitem{lustig_2008}
  Lustig, M., Donoho, D. L., Santos, J. M., and Pauly, J. M. (2008). Compressed
  sensing MRI. IEEE Signal Processing Magazine, 25(2), 72-82.

\bibitem{indyk_1998}
  Indyk, P., and Motwani, R. (1998, May). Approximate nearest neighbors: towards
  removing the curse of dimensionality. In Proceedings of the Thirtieth Annual
  ACM Symposium on Theory of Computing, 604-613.

\bibitem{lin_chu_2018}
  Lin, L. and Chu, H. (2018). Bayesian multivariate meta‐analysis of multiple
  factors. Research Synthesis Methods, 9(2).

\bibitem{olsen_2020}
  Olsen, U., Lindberg, M. F., Denison, E. M. L. et~al.~(2020). Predictors of
  chronic pain and level of physical function in total knee arthroplasty: a
  protocol for a systematic review and meta-analysis. BMJ Open, 10(9), e037674.

\bibitem{riley_2008} 
  Riley R.D., Thompson J.R., and Abrams K.R. (2008). An alternative model for
  bivariate random-effects meta-analysis when the within-study correlations are
  unknown. Biostatistics, 9(1).

\bibitem{riley_2009}
  Riley, R. D. (2009). Multivariate meta‐analysis: the effect of ignoring
  within‐study correlation. Journal of the Royal Statistical Society: Series A
  (Statistics in Society), 172(4), 789-811.

\bibitem{salanti_2011}
  Salanti, G., A. E. Ades, and J. P. Ioannidis (2011), Graphical methods and 
  numerical summaries for presenting results from multiple-treatment 
  meta-analysis: an overview and tutorial. Journal of Clinical Epidemiology, 
  64(2), 163-71.

\bibitem{tange_2020}
  Tange, O. (2020). GNU Parallel 20200722 ('Privacy Shield'). Zenodo, DOI:
  10.5281/zenodo.3956817

\bibitem{thanei_2017}
  Thanei, G. A., Heinze, C., and Meinshausen, N. (2017). Random projections for
  large-scale regression. In Big and Complex Data Analysis, 51-68, 
  Springer.

\bibitem{white_2009}
  White IR (2009). Multivariate random-effects meta-analysis. Stata Journal; 9,
  40-56.

\bibitem{white_2011}
  White IR (2011). Multivariate random-effects meta-regression: Updates to mvmeta.
  Stata Journal; 11, 255-270.

\bibitem{white_2015}
  White, I. R. (2015). Network meta-analysis. Stata Journal, 15(4), 951-985.

\end{thebibliography}

\section*{Author contributions}

CJR developed the model, performed the analyses, and wrote the manuscript. UO,
MFL, EMLD, AA, and AL planned the systematic review, collected data, and
contributed to the manuscript.

\section*{Funding}

This work is supported by funding from the Norwegian Research Council of Norway
(grant \#287816) and the South-Eastern Regional Health Authority (grant
\#2018060).

\section*{Conflicts of interest}

Within the previous five years, CJR was employed by OncoImmunity AS. He has
patents and patent applications with no relevance to this study. The other authors
do not report any potential conflicts of interest.

\end{document}